\documentclass[prd,aps,reprint,nofootinbib,showpacs,superscriptaddress]{revtex4-1}
\usepackage{graphicx} 
\usepackage{hyperref}
\usepackage{amsfonts}
\usepackage{amsmath,amssymb}
\usepackage{bm} 
\usepackage{color}
\usepackage{epstopdf} 
\usepackage{epsfig}
\usepackage{subfig} 
\usepackage{float}

\def\be{\begin{equation}} 
\def\ee{\end{equation}}  
\def\bea{\begin{eqnarray}}
\def\eea{\end{eqnarray}}
\def\ba{\begin{array}}
\def\ea{\end{array}}
\def\bem{\begin{multline}} 
\def\eem{\end{multline}}

\begin{document}

\title{Superconducting Dome from Holography}

\author{Suman Ganguli}
\email{sgangul3@tennessee.edu}
\affiliation{Department of Physics and Astronomy, The University of Tennessee, Knoxville, Tennessee 37996-1200}
\author{Jimmy A. Hutasoit}
\email{jah77@psu.edu} 
\affiliation{Department of Physics, The Pennsylvania State University, University Park, Pennsylvania 16802}
\affiliation{Department of Physics, West Virginia University, Morgantown, West Virginia 26506}
\author{George Siopsis}
\email{siopsis@tennessee.edu}
\affiliation{Department of Physics and Astronomy, The University of Tennessee, Knoxville, Tennessee 37996-1200}

\date{\today} 

\begin{abstract}
We find a regime in which a strongly coupled striped superconductor features a superconducting dome. This regime is signified by i) a modulating chemical potential that averages to zero, and ii) a superconducting order parameter that has a scaling dimension $3/2 < \Delta \le 3$. We also find that in this regime, the order parameter exhibits a mild dependence on the modulation wavelength of the stripes.
\end{abstract}

\maketitle

\section{Introduction}

Quantum many-body physics lacks a general mathematical framework that can deal with strongly coupled fermions at finite density. In an effort to remedy this problem, recently, there has been a flurry of activities applying gauge/gravity duality to study strongly coupled condensed matter systems (for a review, see, e.g., \cite{Sachdev:2011fk}). A large portion of these activities revolves around understanding high temperature (high $T_c$) superconductors, such as cuprates and iron pnictides. One of the many challenges in adopting this approach is to reproduce the phase diagrams of the high $T_c$ superconductors. The problem is that it is not clear how one should introduce the effect of doping into the holographic model. 

Being strongly coupled, the normal states of high $T_c$ superconductors are highly correlated, and thus exhibit other low temperature orders which interact with superconductivity in an intricate way. One of the prominent orders is the unidirectional charge density wave or ``stripe" order \cite{Hoffman:2002,Howald:2003,Vershinin:2004,Wise:2008}. However, the complication that comes from the existence of these coexisting orders might also be the key toward reconstructing the phase diagrams of high temperature superconductors. In particular, a very important hint comes from the observation that in the hole doped cuprates, in particular Bi$_{2-y}$Pb$_y$Sr$_{2-z}$La$_z$CuO$_{6+x}$, near the superconducting regime the modulation wavenumber is a monotonically increasing function of doping \cite{Beyer2008471,Wise:2008}, which is presumably valid in both the underdoped and overdoped regimes. Because of this, the qualitative behavior of the critical temperature viewed as a function of either doping or the modulation wavenumber should be similar. Therefore, as a step toward realizing a realistic holographic model of cuprates, it is crucial to show that indeed there exists a regime in the parameter space in which the holographic striped superconductor exhibits a superconducting dome.

\section{Set-up}

In order to achieve the goal set in the previous section, following \cite{Ganguli:2012fk}, we consider an Einstein-Maxwell-scalar system on a $3+1$-dimensional spacetime with a negative cosmological constant $\Lambda = -3/L^2$ as a holographic model of a $2+1$-dimensional strongly coupled striped superconductor. The bulk scalar field is dual to the superconducting order parameter, while the bulk $U(1)$ gauge field is dual to the four-current in the strongly coupled system. For simplicity, we shall adopt units in which $L=1$, $16\pi G =1$. We introduce the stripe order phenomenologically by introducing a modulating chemical potential,
\be \mu(x) = \mu(1-\delta) + \mu \delta \cos Qx~, \ee
which is identified as the boundary value of the bulk electrostatic potential. We are interested in a black hole solution with this boundary condition and the superconducting transition temperature will then be given by the critical temperature below which the black hole forms scalar hair. 

It has been shown that if the average value of the chemical potential does not vanish ($\delta \ne 1$), then the critical temperature will have a long power law tail at large $Q$. This is true both at the mean field level\footnote{This is equivalent to the probe limit in the gravity picture, where the backreaction of the $U(1)$ field and the scalar field on the spacetime geometry is neglected.} \cite{Hutasoit:2012fk} and when fluctuations are turned on\footnote{Equivalently, when backreaction is included.} \cite{Ganguli:2012fk}. A similar result is also obtained when $\delta=1$, and the scaling dimension of the order parameter satisfies $1/2 < \Delta \le 3/2$. This is an unwanted feature since to obtain a superconducting dome, we need to have a critical modulation wavenumber $Q_{\ast}>0$, above which the critical temperature vanishes. At the mean field level, it was shown in Ref.  \cite{Hutasoit:2012fk} that the $\delta=1$ and $3/2<\Delta\leq3$ case results in such a feature.

Another hint we obtained from our previous work is that in the $\delta \ne 1$, and the $\delta=1$, $1/2 < \Delta\le 3/2$ cases, fluctuations cause a significant drop in the critical temperature at small $Q$ \cite{Ganguli:2012fk}. Combining the above, we expect that when the chemical potential averages to zero ($\delta=1$) and the scaling dimension of the order parameter satisfies $3/2<\Delta\leq3$, the superconducting regime is capped at larger values of $Q$, while when the fluctuations are turned on, it is capped at the smaller values of $Q$, thus resulting in a superconducting dome. We will show that indeed this expectation is correct.

The equations of motion consist of the Einstein equations
\be R_{ab} - \frac{1}{2} g_{ab} R - 3g_{ab} = \frac{1}{2} T_{ab}~,\ee
the Maxwell equations
\be \frac{1}{\sqrt{-g}} \partial_{b}(\sqrt{-g}F^{ab}) = J^{a}~,\ee
and the Klein-Gordon equation for the scalar field
\be\label{eq4} -D_{a}(\sqrt{-g}g^{ab}D_{b}\phi)/\sqrt{-g}+m^2 \phi=0~.\ee
Here, $T_{ab}$ is the stress-energy tensor,
the $U(1)$ current is $J^{a}=-i[\phi^* D^{a}\phi- \mathrm{c.c.} ]$,
the covariant derivative is $D_a=\partial_a-iqA_a$, the $U(1)$ field strength is $F_{ab}=\partial_{a} A_{b}-\partial_{b} A_{a}$ and $a,b\in\{ t,z,x,y\}$. Furthermore, $q$ is the charge of the scalar field, which is related to the central charge of the conformal field theory describing the critical point and $m$ is the mass, which is related to the scaling dimension $\Delta$ by $m^2 = \Delta (\Delta-3)$.

\section{Superconducting dome}

We solve the equations of motion by doing an expansion in $1/q^2$, where taking $q\to \infty$ while keeping $q\mu$ fixed corresponds to the probe limit. In the normal state, where the order parameter vanishes, and therefore $\phi=0$, the solution for the metric for arbitrary $\delta$ is given in \cite{Ganguli:2012fk}. Up to ${\cal O}(1/q^2)$, the solution for $\delta = 1$ is then given by
\be
ds^2 = \frac{r_+^2}{z^2} \left[ - h\left(1-\frac{\alpha}{q^2}\right) dt^2 + \frac{1+ \alpha/q^2}{r_+^2h} dz^2 + d\vec{x}^2\right],
\ee
where the boundary is at $z=0$ and the horizon is at $z=1$. Here, $h = 1-z^3$ and $\alpha = \alpha^{(0)}(z) + \alpha^{(2)}(z) \cos 2Qx$, where
\be
\alpha^{(0)} = \frac{z^3}{8h} \int_z^1 \left[\left(\partial_{z'}{\cal A}_0\right)^2 - \frac{Q^2{\cal A}_0^2}{r_+^2h}\right]dz',
\ee
and $\alpha^{(2)}$ satisfies
\bea
{\alpha^{(2)}}'-\left(\frac{3 }{z }-\frac{ 2Q^2 z+r_+^2h'}{r_+^2 h}\right) \alpha^{(2)} & & \nonumber \\
+\, \frac{z^3 \left(Q^2 \mathcal{A}_0^2+ r_+^2h {\mathcal{A}_0'}^2\right)}{8 r_+^2 h^2} &=& 0~.
\eea
Prime $'$ denotes a derivative with respect to z. The temperature is given by
\be
T = \frac{3 r_+}{4\pi } \left[ 1 - \frac{1}{q^2}\frac{ {\mathcal{A}_0'}^2 }{24} \bigg|_{z=1} \right].
\ee
The electrostatic potential is given by $A_t = \left({\cal A}_0 + {\cal A}_1/q^2 \right) \cos Qx$, where ${\cal A}_0$ and ${\cal A}_1$ satisfy
\bea
{{\cal A}_0}'' - \frac{Q^2}{r_+^2h} {\cal A}_0=0, \qquad {{\cal A}_1}'' - \frac{Q^2}{r_+^2h} {\cal A}_1= \frac{Q^2}{r_+^2h}  \alpha^{(0)} {\cal A}_0. \nonumber \\
\eea

\begin{figure}[H]
\begin{center} 
\includegraphics[width=3.3 in]{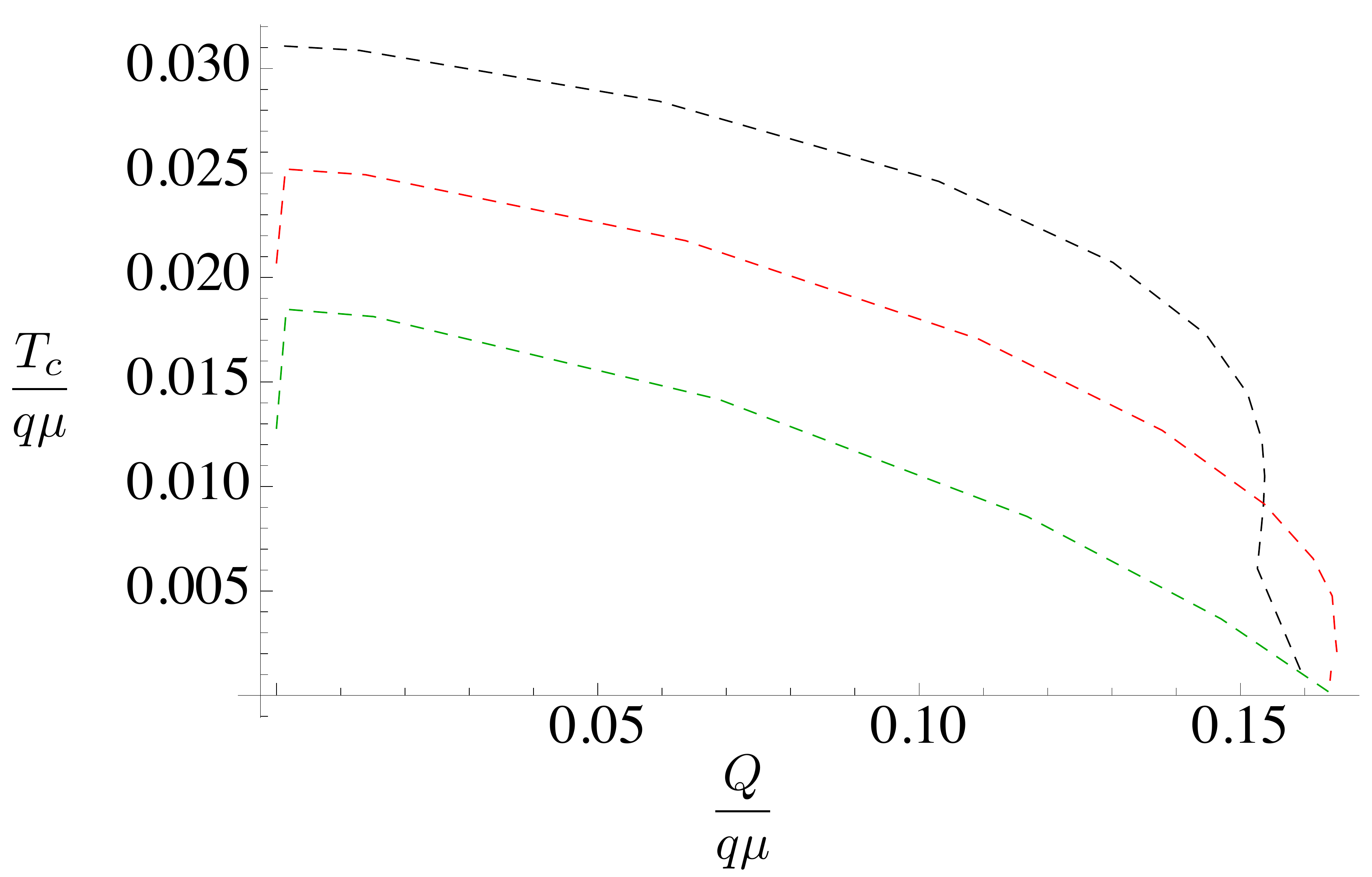}
\end{center}  
\caption{(Color online) The critical temperature as a function of modulation wavenumber. Here, $\Delta =3$ and the black, red and green lines correspond to $q^2 = \infty$, 10 and 5, respectively. \label{Tc} }
\end{figure}

As we lower the temperature, $\phi=0$ becomes an unstable solution to the scalar equation of motion \eqref{eq4}, which takes the form
\begin{multline}\label{Scalar PDE}  
\sum_{i=z,x} \frac{1}{\sqrt{-g}}\, \partial_i \left( \sqrt{-g} g^{ii} \partial_i \phi \right) +  \left( q^2 g^{tt}\, A_t^2  - m^2 \right) \phi =0\,.
\end{multline} 
To solve this, we expand $\phi$ in a Fourier series,
\bea\label{DefPsiF}
\phi(z,x) &=& \frac{\langle {\cal O}_{\Delta} \rangle z^\Delta}{\sqrt{2}r_+^{\Delta}} \, \sum F^{(n)}(z) \cos n Q x ~,
\eea
where $\langle {\cal O}_{\Delta} \rangle$ is the superconducting order parameter. We evaluate the critical temperature numerically and a typical result can be seen in Fig. \ref{Tc}.

Turning on fluctuations is equivalent to decreasing the value of $q^2$ and this results in the critical temperature decreasing at small $Q$. The discontinuous behavior exhibited at $Q \to 0$ is an artifact of the first-order contribution to the perturbative expansion of the solutions of the Einstein-Maxwell equations.
We expect that as we include higher orders in perturbation, the graph will become smoother. 

In the regime of validity of the $1/q^2$ expansion, we are not able to crank up the fluctuations to reach vanishing critical temperature at $Q=0$. However, as the exact solution at $Q=0$ is known, we expect this to happen at \cite{Horowitz:2009fk}
\be q^2 = \frac{3}{4}+\frac{\Delta (\Delta-3)}{2}~. \ee
It would be interesting to extend our calculations down to zero temperature. This would require going beyond first order in the perturbative expansion. To this end, it may be advantageous to apply new numerical techniques such as those that have been recently developed in Ref. \cite{Horowitz:2013fk}.

Unlike the cases considered in Ref. \cite{Ganguli:2012fk}, here we see that there is a critical modulation wavenumber $Q_{\ast}$, above which the critical temperature vanishes. The value $Q_{\ast}$ can be estimated analytically by considering the large modulation wavenumber regime where $Q\gg T_c$. In this regime, we can see from the numerics that the effects of backreaction are suppressed. Furthermore, the higher modes are suppressed. This behavior should be contrasted with the behavior of the homogeneous system ($Q=0$). In that case, the critical temperature is lowered from its probe limit value by decreasing $q$ and backreaction plays a crucial role. As one increases $Q$, the effects of the backreaction decrease, and for a fixed $q$ the critical temperature decreases, eventually reaching zero. Thus zero critical temperature can be approached by keeping $Q$ fixed and decreasing $q$ (i.e., increasing backreaction), or by keeping $q$ (i.e., backreaction) fixed and increasing $Q$.

For an explicit calculation of the critical value $Q_\ast$, using the perturbative method of Ref.  \cite{Hutasoit:2012fk}, we have
\be 
\frac{r_{+c}^2}{q^2\mu^2} =\frac{1}{2\Delta-3} \left(\tilde{a}_c(1) - \frac{\Gamma^2\left(\tfrac{\Delta}{3}\right)}{\Gamma\left(\tfrac{2\Delta}{3}\right)} \, \frac{\Gamma\left(\tfrac{2(3-\Delta)}{3}\right)}{\Gamma^2\left(\tfrac{3-\Delta}{3}\right)} a_c(1)\right), \label{eq:r+c}
\ee
where 
\bea
a_c(1) &=& \frac{1}{2^{2\Delta}}\frac{r_{+c}^{2\Delta -1}}{Q^{2\Delta -1}}  \Gamma(2\Delta -1), \nonumber\\ 
\tilde{a}_c(1) &=& \frac{r_{+c}^2}{8 Q^2}.
\eea
Here, $r_{+c}$ is the value of $r_+$ at the critical temperature. Therefore,
\bea
\frac{r_{+c}^{2\Delta-3}}{(q \mu)^{2\Delta-3}} &=& \frac{2^{2\Delta}(2\Delta-3)}{\Gamma(2\Delta-1)} \frac{\Gamma\left(\tfrac{2\Delta}{3}\right)}{\Gamma^2\left(\tfrac{\Delta}{3}\right)} \, \frac{\Gamma^2\left(\tfrac{3-\Delta}{3}\right)}{\Gamma\left(\tfrac{2(3-\Delta)}{3}\right)} \frac{Q^{2\Delta-1}}{(q \mu)^{2\Delta-1}} \nonumber \\
& & \times \left(\frac{1}{8(2\Delta-3)} \frac{q^2 \mu^2}{Q^2} - 1 \right),
\eea
which means that above 
\be
\left(\frac{Q_{\ast}}{q \mu}\right)^2 = \frac{1}{8(2 \Delta-3)}, \label{eq:qastzero}
\ee
we have no instability and $T_c=0$.

\begin{figure}
\begin{center} 
\includegraphics[width=3.4 in]{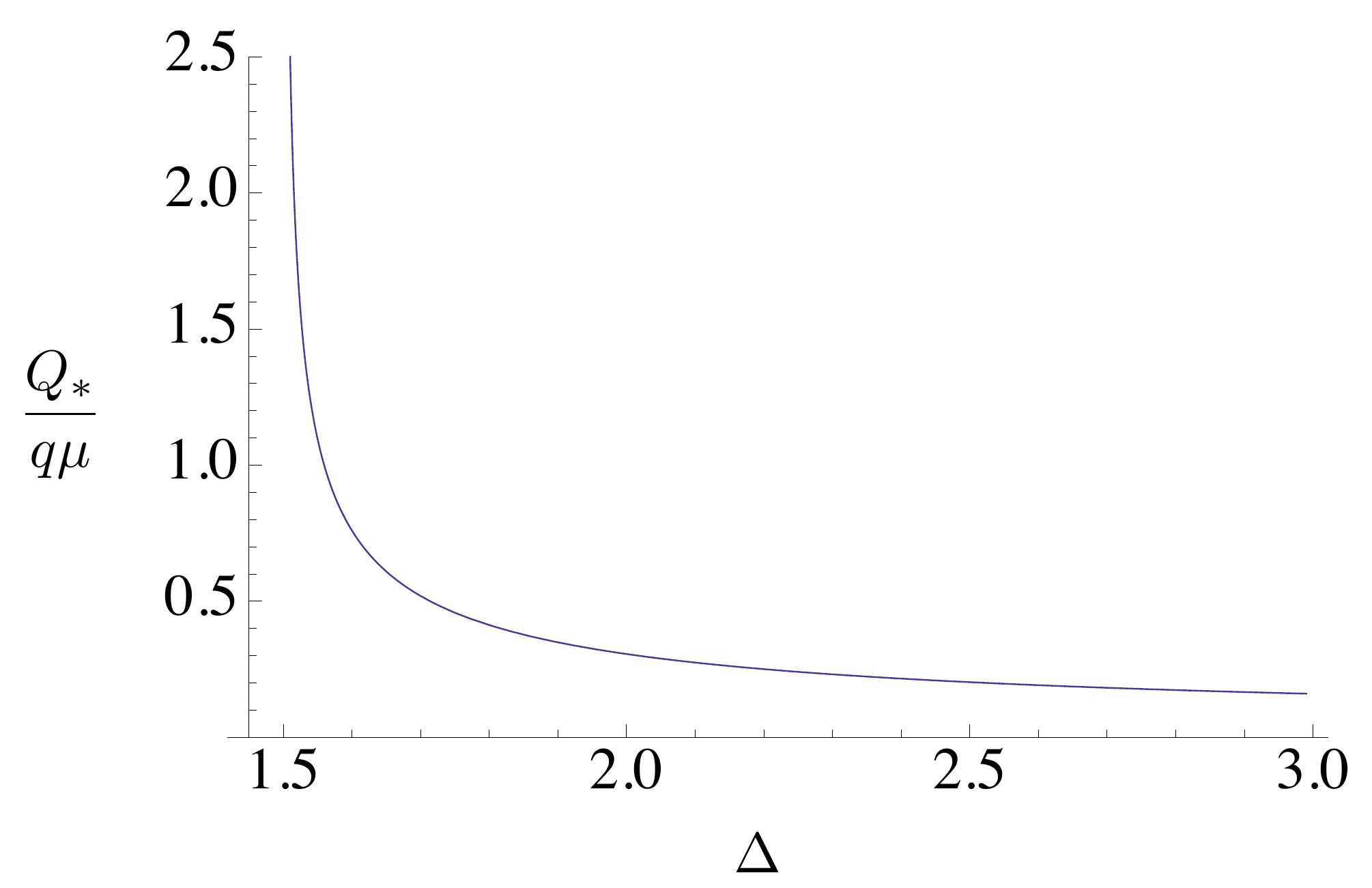}
\end{center}  
\caption{The end point of the superconducting dome as a function of the scaling dimension of the order parameter. \label{fig:Qast} }
\end{figure}

We can improve upon this approximation by iteratively solving the equation of motion 
\be\label{eq0} 
\partial^2_{\tilde{z}}F^{(0)} + \frac{2(\Delta -1)}{\tilde{z}} \partial_{\tilde{z}}F^{(0)} + \lambda e^{-\tilde{z}} F^{(0)} = 0, 
\ee
where $\tilde{z} = zr_+/(2Q)$ and $\lambda = q^2 \mu^2/(8 Q_{\ast}^2)$. We would like to evaluate this in the interval $\tilde{z} \in [0,\infty)$, with the boundary condition $F^{(0)} \to 0$ as $\tilde{z} \to \infty$. This differential equation can be solved exactly for $\Delta = 2$, but for other values $3/2<\Delta\leq3$, we can estimate $\lambda$ by iteration. To do so, let us rewrite Eq. \ref{eq0} as an integral equation
\bea\label{inteq} 
F^{(0)} (\tilde{z}) &=& 1 - \frac{\lambda}{2\Delta -3}\int_0^{\tilde{z}} dw\, w\,  e^{-w}\, F^{(0)}(w) \nonumber \\
& & +\, \frac{\lambda}{(2\Delta -3)\tilde{z}^{2\Delta -3}} \int_0^{\tilde{z}} dw \, w^{2\Delta -2} e^{-w} \,F^{(0)}(w). \nonumber \\
\eea
Solving this iteratively, we reproduce Eq. \ref{eq:qastzero} at zeroth order while a first order correction results in
\bea
\left(\frac{Q_{\ast}}{q \mu}\right)^2 &=& \frac{3-4\Delta}{32} + \frac{(\Delta -1)(\Delta -2)}{4}  \times\nonumber\\
& & \times \left[ \psi \left(\frac{5}{2} -\Delta\right) - \psi (3-\Delta) + \frac{2\pi}{\sin 2\pi\Delta} \right]. \label{eq:endpoint}\nonumber\\
\eea
Here, $\psi$ denotes the digamma function. Despite appearances, as we can see in Fig. \ref{fig:Qast}, this is a smooth function of $\Delta$ in the interval $(3/2,3)$ (the two expressions above have almost indistinguishable graphs). Furthermore, at $\Delta \to 3^-$, we have $\frac{Q_{\ast}}{q \mu} \approx 0.158$, in good agreement with our numerical results.

By comparing the information concerning the end points of the dome, $Q=0$ and $Q=Q_{\ast}$, with experiment, we can then in principle extract the value of $\Delta$. This is because the $Q=0$ point gives us the information on $q$ and upon substituting it into Eq. \ref{eq:endpoint}, we obtain $\Delta$.

Another method to extract the realistic value of $\Delta$ is by calculating the anisotropy of the optical conductivity akin to the calculation done in \cite{Hutasoit:2012vn} (see also \cite{Horowitz:2012fk}) and comparing its scaling behavior with the observed behavior in cuprates \cite{Lee:2005fk}. Work in this direction is in progress.

\section{Gap behavior}

\begin{figure}
\begin{center} 
\includegraphics[width=3.5 in]{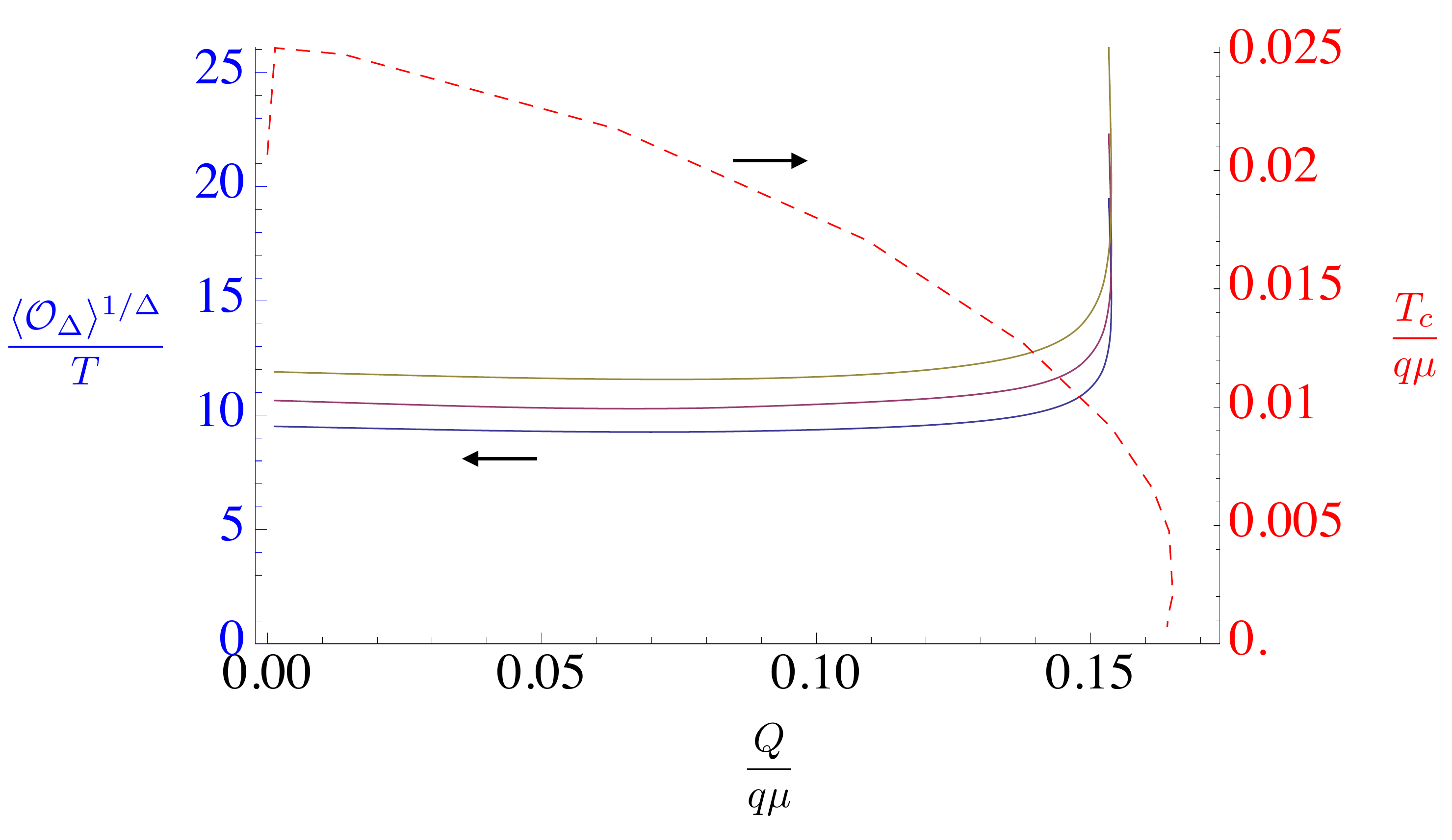}
\end{center}  
\caption{(Color online) The ratio of the gap with respect to temperature as a function of the wavenumber at fixed $T/T_c$. Here, $\Delta=3$ and the blue, purple and yellow lines (bottom to top) correspond to $T/T_c = 0.975$, 0.95 and 0.9, respectively. For clarity, we have also shown the critical temperature, plotted with a dashed line. \label{fig:belowTc} }
\end{figure}

Now that we have understood the behavior of the system at the critical temperature, we can go to a temperature $T$ below $T_c$ and study the behavior of the order parameter. Limiting ourselves to the case where the bulk scalar field can be considered as a small perturbation to the solution at critical temperature, we obtain the gap at fixed $T/T_c$ as plotted in Fig. \ref{fig:belowTc}. This corresponds to the temperature below but near $T_c$.

We see that deep under the superconducting dome, the gap shows a mild dependence on $Q$. Interestingly, measurement of the gap at fixed low temperatures also showed mild dependence on $Q$ \cite{Wise:unp}. We are hopeful that gap measurements near $T_c$ from STM will be available to be compared with our prediction in Fig. \ref{fig:belowTc} in the near future. 

\begin{figure}
\begin{center} 
\includegraphics[width=3.5 in]{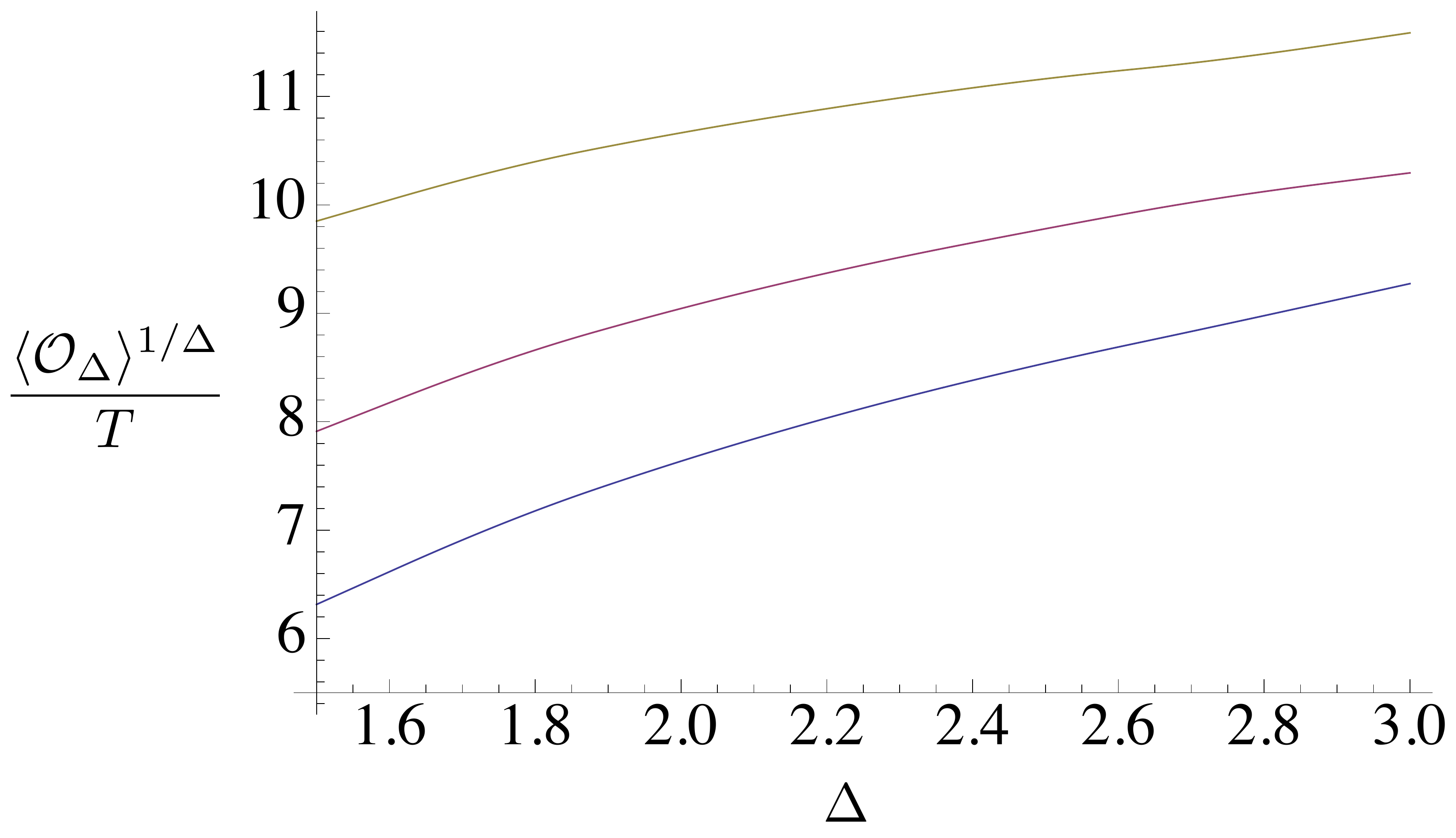}
\end{center}  
\caption{(Color Online) The ratio of the gap to temperature deep under the dome, as a function of the scaling dimension of the order parameter. Here,  the blue, purple and yellow lines (bottom to top) correspond to $T/T_c = 0.975$, 0.95 and 0.9, respectively. \label{fig:gapDelta}}
\end{figure}

The ratio $\langle {\cal O}_{\Delta} \rangle^{1/\Delta}/T$ deep under the superconducting dome as a function of $\Delta$ is plotted in Fig. \ref{fig:gapDelta}. The value is about 8--10, which is the same order of magnitude as the measured value of 29 \cite{Wise:2008}. We would like to note that the measurement is done at $T/T_c \approx 0.2$, which is beyond the regime of validity of our approximation. However, considering that in some holographic models, the ratio of gap over temperature can go as high as ${\cal O} (10^3)$ (see for example \cite{Alsup:2011fk}), our result is encouraging.

The result at the far end of the dome $Q \approx Q_{\ast}$ can be understood analytically by again using the perturbative method of Ref. \cite{Hutasoit:2012fk}. We have
\be
\frac{r_+^2}{q^2\mu^2} =\frac{1}{2\Delta-3} \left(\tilde{a}(1) - \frac{\Gamma^2\left(\tfrac{\Delta}{3}\right)}{\Gamma\left(\tfrac{2\Delta}{3}\right)} \, \frac{\Gamma\left(\tfrac{2(3-\Delta)}{3}\right)}{\Gamma^2\left(\tfrac{3-\Delta}{3}\right)} a(1)\right) \label{eq:1},
\ee
where
\bea
a(z) &=& \int^z_0 \frac{dz'}{{z'}^{2-2 \Delta}} F^{(0)} \, \frac{{\cal A}_0}{h} \, {F}^{(0)}, \nonumber \\
\tilde{a}(z) &=& \int^z_0 \frac{dz'}{{z'}^{2-2 \Delta}} \tilde{F}^{(0)} \, \frac{{\cal A}_0}{h} \, {F}^{(0)}.
\eea
Here,
\bea
{F^{(0)}} &=& \,_2F_1 \left(\frac{\Delta}{3}, \frac{\Delta}{3}; \frac{2 \Delta}{3}; z^3 \right),\\
{\tilde{F}^{(0)}} &=& z^{3 - 2 \Delta} \, \,_2F_1 \left(\frac{3 -\Delta}{3}, \frac{3 - \Delta}{3}; \frac{2 (3 -\Delta)}{3}; z^3 \right), \nonumber
\eea
and
\be
{\cal A}_0 = \frac{e^{-2Qz/r_+}}{2} \, \left(1 -  \frac{z^{2\Delta-1}}{(2 \Delta-1)Q} \,  \frac{{\langle {\cal O}_{\Delta} \rangle^{(0)}}^2}{r_+^{2\Delta -1}}\right). \label{eq:el below Tc}
\ee
We note that by setting the order parameter to zero in Eq. \ref{eq:el below Tc}, we obtain the value for $r_{+c}^2/(q^2\mu^2)$ from Eq. \ref{eq:1}, which is given by Eq. \ref{eq:r+c}. Then, by dividing Eq. \ref{eq:1} with Eq. \ref{eq:r+c}, after some algebra we obtain
\bea
\frac{{\langle {\cal O}_{\Delta} \rangle^{(0)}}^2}{T_c^{2\Delta}} &=& \frac{2}{\Delta} \left(\frac{4 \pi}{3}\right)^{2\Delta-3}\frac{\Gamma^2\left(\tfrac{\Delta}{3}\right)}{\Gamma\left(\tfrac{2\Delta}{3}\right)} \, \frac{\Gamma\left(\tfrac{2(3-\Delta)}{3}\right)}{\Gamma^2\left(\tfrac{3-\Delta}{3}\right)} \nonumber \\
&& \times \left(\frac{Q}{T_c} \right)^{3} \left[1-\left(\frac{T}{T_c}\right)^{2\Delta-3}\right]. 
\eea
We see that as $Q\to Q_{\ast}$ that even though the gap vanishes, the ratio of the gap to critical temperature diverges as $T_c^{-3/(2\Delta)}$. It is interesting that we found a regime in which the value of $\langle {\cal O}_{\Delta} \rangle^{1/\Delta}/T_c$ gets very large at low temperature. In comparison, the values for homogeneous BCS superconductors and homogeneous holographic superconductors are 3.54 and 8, respectively \cite{Alsup:2011fk}.

\section{Summary and Discussions}

In this article, we have shown that when the homogeneous part of the chemical potential vanishes and when the scaling dimension of the order parameter is larger than 3/2, the holographic striped superconductor exhibits a superconducting dome with vanishing or nearly vanishing critical temperature at the end points. Fitting the features of the end points to experimental data will give us the central charge of the conformal field theory describing the critical point and the scaling dimension of the superconducting order parameter.

We have also shown that deep under the superconducting dome, the order parameter exhibits a mild dependence on the modulation wavenumber. This is perhaps somewhat unexpected and shows that the lack of $Q$-dependent features in the superconducting gap data should not be taken as an indication that the stripe ordering is not related to the mechanism that gives rise to superconductivity.

Toward the end of the superconducting dome, the gap exhibits qualitatively different behavior. In particular, as one approaches $Q=Q_{\ast}$, the gap decreases to zero while its ratio with respect to critical temperature increases.

It is important that we remind the reader that in this holographic model, the stripe order is introduced phenomenologically in the form of a modulated chemical potential. In order to understand the phase diagram of the holographic striped superconductor more accurately, we ultimately would like to have both the stripe order and superconducting order emerging dynamically\footnote{Holographic models for stripe formation have been introduced in Refs. \cite{Nakamura:2009fk,Donos:2011uq,Alsup:2012fk,Iizuka:2013fk}.}.  Work in this direction is in progress.

Lastly, let us make a comment comparing superconducting domes of hole doped cuprates and heavy fermion systems such as CePd$_2$Si$_2$. In heavy fermion systems, the existence of a superconducting dome is tightly related to the existence of a magnetic quantum critical point due to the belief that magnetic fluctuations play a critical role in the superconducting mechanism \cite{Monthoux:2007fk}. It is tempting to imagine that perhaps similar physics arises in the hole doped cuprates, where instead of magnetic ordering, one has charge/stripe ordering. However, even though some holographic models of stripe formation introduced so far exhibit instabilities at zero temperature, it is not clear whether these are related to quantum critical points.

\acknowledgments 
We would like to thank Eric Hudson for valuable discussions. The work of S.\ G.\ and G.\ S.\ is supported in part by the Department of Energy under grant DE-FG05-91ER40627. J.\ H.\ is supported by West Virginia University start-up funds at the early stage of this work and by NSF grant DMR-1005536 and DMR-0820404 (Penn State MRSEC) at the later stage.  

\bibliography{References} 
\end{document}